\documentclass[aip, apl, reprint]{revtex4-1}
\usepackage[utf8x]{inputenc}
\usepackage{graphicx}
\usepackage{amsmath}
\usepackage{natbib}
\usepackage[caption=false]{subfig}

\begin{document}

\title{
Minimal graphene thickness for wear protection of diamond}
\author{M. M. van Wijk}
\affiliation {Radboud University Nijmegen, Institute for Molecules and Materials, Heyendaalseweg 135, 6525 AJ Nijmegen, The Netherlands}
\author{A. Fasolino}
\affiliation {Radboud University Nijmegen, Institute for Molecules and Materials, Heyendaalseweg 135, 6525 AJ Nijmegen, The Netherlands}

\begin{abstract}
We show by means of molecular dynamics simulations that graphene is an excellent coating for diamond. The transformation of diamond to amorphous carbon while sliding under pressure can be prevented by having at least two graphene layers between the diamond slabs, making this combination of materials suitable for new coatings and micro- and nanoelectromechanical devices. Grain boundaries, vacancies and adatoms on the diamond surface do not change this picture whereas reactive adsorbates between the graphene layers may have detrimental effects. Our findings can be explained by the properties of layered materials where the weak interlayer bonding evolves to a strong interlayer repulsion under pressure.
\end{abstract}

\maketitle

Control of friction and wear is one of the key challenges for the design of micro- and nanoelectromechanical systems (MEMS/NEMS). There is an ongoing quest to make these devices reliable, robust and able to resist demanding environments, under high stress and with sliding surfaces in contact. 
For  these devices, lubrication has to be based on dry solid coatings rather than on liquids to avoid undesirable effects, associated with viscosity~\cite{kim2007nanotribology}, squeeze out~\cite{persson2004squeeze} and stiction~\cite{bhushan2010springer}.

The present MEMS/NEMS technology is based on silicon~\cite{sumant2010ultrananocrystalline, berman2013surface}, but its poor mechanical, chemical and tribological properties make alternatives desirable and actively sought after~\cite{guisinger2010beyond}.
In particular, at the nanoscale wear is a limiting factor as it drastically shortens their lifetime~\cite{bhushan2010springer}.
Diamond is such an alternative material in view of its hardness and chemical inertness. Making perfectly crystalline diamond is difficult, but nanocrystalline diamond (grain sizes of 10-200~nm) shares many of its properties and is attainable by CVD~\cite{sumant2010ultrananocrystalline}.

Although diamond is very hard, it is not resistant to wear and it can be polished. The polishing rate has been shown to depend on the surface orientation and sliding direction~\cite{pastewka2010anisotropic}. The amorphous layer which develops at the sliding interface is easily removed leading to wear of the surface. This amorphous phase, with many bonds at the interface,
leads to a high friction coefficient. Fortunately, lowering of the friction coefficient after some time, also called running-in, is observed for sliding amorphous carbon~\cite{fontaine2005achieving, pastewka2010atomistic}.
The microscopic mechanisms for this behavior are still a matter of debate. Molecular dynamics based on a modified version of the empirical potential REBO, reports the formation of a graphene-like layer~\cite{ma2014shear} during sliding under pressure that would inhibit the further growth of an amorphous layer at the interface. Recent ab-initio calculations, instead, attribute the reduction of friction after the initial phase (running in)
to passivation of the dangling bonds by water or, preferably, by hydrogen~\cite{konicek2008origin,righi2011abinitio,debarros2012friction}. For the latter, a minimum humidity or hydrogen gas pressure is necessary and the contact pressure needs to be below a critical value~\cite{konicek2008origin} for passivation.
These results suggest that operation in vacuum or high-pressure environments would be difficult. 

An approach to reduce wear is to look for suitable coatings, effective at the nanoscale. Moreover, it  is desirable to have a very thin coating. 
Graphene is a natural candidate for this purpose in view of its exceptional mechanical properties~\cite{young2012mechanics}. The frictional properties of (few-layer) graphene have been recently intensively studied~\cite{filleter2009friction,lee2010frictional} showing a lowering of friction with decreasing number of layers. 
Coating of sliding steel surfaces with few layer graphene has been demonstrated to reduce drastically the friction and wear during sliding~\cite{berman2012few}. On a smaller scale, coating an AFM probe with graphene also improved resistance to wear~\cite{martin2013graphene}. Moreover it has been recently shown that graphene withstands without damage much higher  loads than diamond-like carbon~\cite{sandoz2012atomistic}, making graphene suitable for high-pressure conditions.

Here we suggest combining the properties of diamond and graphene to form a hard but smoothly sliding structure to enable new MEMS/NEMS technologies. 

We perform atomistic simulations to describe the wear of diamond surfaces during sliding under pressure when the surfaces are  either bare or  separated by one or two layers of graphene.
We find that at least two layers of graphene form a contact that drastically reduces friction and wear.  

The interatomic interactions are given by the reactive empirical potential LCBOP~\cite{los2003intrinsic}  as implemented in the molecular dynamics code LAMMPS~\cite{lammps}. This bond-order potential can accurately describe different phases of carbon~\cite{lucaprl}, the transformations between them and the elastic constants of diamond and graphite. It can also describe the interaction of single carbon atoms with the diamond surfaces and graphene. Since single atoms are very reactive we use them to represent the effect of reactive species and impurities. 

\begin{figure}[htbp]
 \centering
 \includegraphics[width=\linewidth]{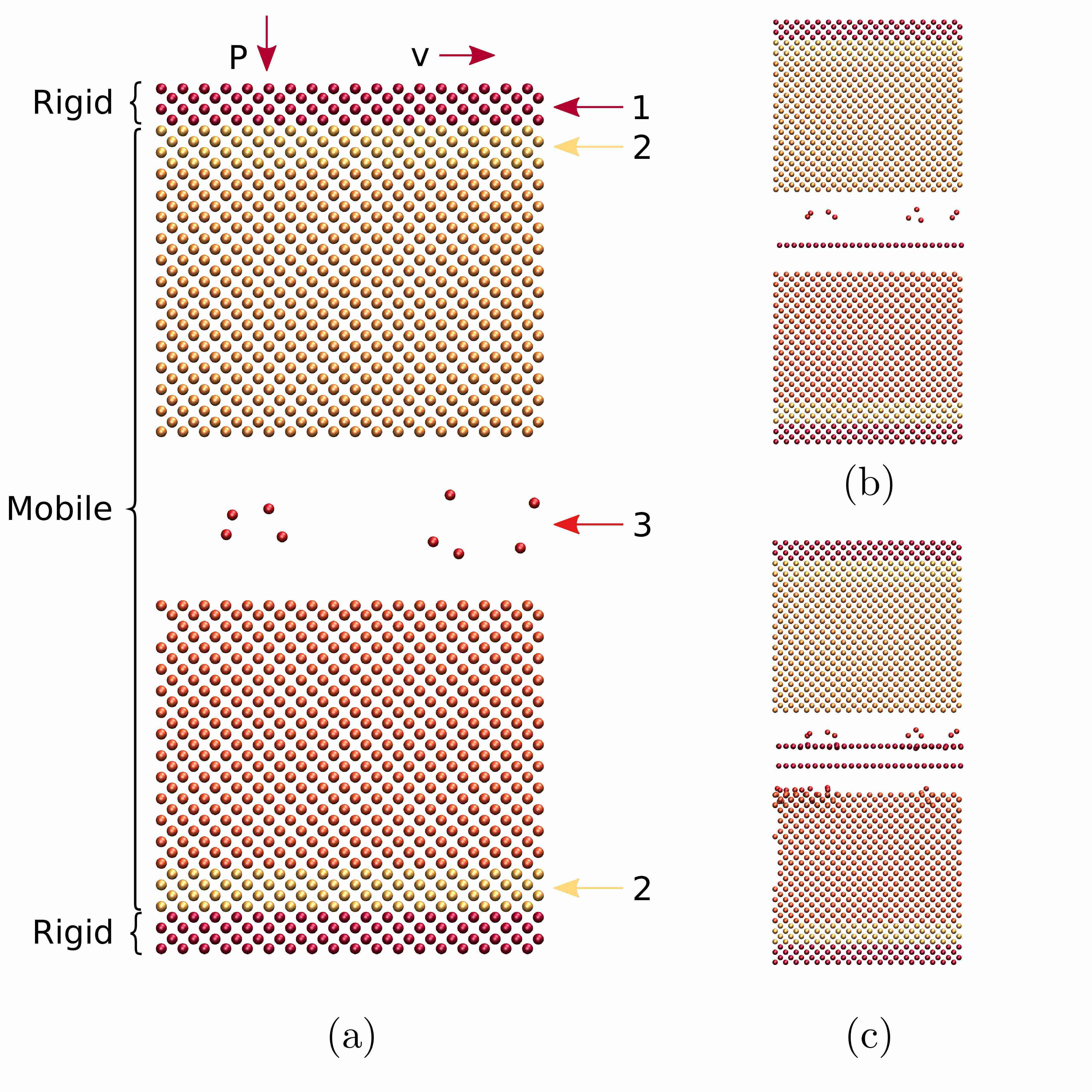}
 \caption{(Color online) (a) Model of the simulated system, consisting of two diamond slabs, each formed by 3564 atoms (108 per atomic layer). We use periodic boundary conditions in the in-plane $x$ and $y$ direction. The top and bottom 4 atomic layers (red, 1) are kept rigid. The top rigid part moves with a constant velocity in the $x$ direction and can move in the $z$ direction as a consequence of the applied pressure and interactions with the mobile atoms. The temperature is controlled by a Langevin thermostat applied to four atomic layers adjacent to the rigid parts (yellow, 2). A few randomly placed atoms (3) prevent cold welding of the slabs. (b) Initial configuration with one layer of graphene of 260 atoms between the sliding diamond surfaces. (c) Initial configuration with two layers of graphene.}
 \label{fig:Setup}
\end{figure}

In Fig.~\ref{fig:Setup} we show a sketch of our model. Our initial sample consists of two slabs of diamond with  $\left(100\right)$ surfaces, which are pressed against each other. The $\left(100\right)$ surface has a square unit cell given by one face of the cubic lattice with lattice parameter 3.5668~\AA. Periodic boundary conditions are imposed in the in-plane $x$ and $y$ directions. Each diamond slab is made of 9 $\times$ 6 $\times$ 8 unit cells. This size is chosen to avoid strain and match the periodic boundary conditions when one or two graphene layers of 260 atoms each are placed between the diamond slabs, as shown in Fig.~\ref{fig:Setup}b,c. The top and bottom 4 atomic layers are kept rigid. The bottom rigid part is kept still, whereas the top rigid part moves in the $x$ direction ($\left<100\right>$ direction) at a fixed velocity $v$=30~m/s. The top rigid part can also move as a whole in the $z$ direction under the influence of a constant force on each atom, which results in a pressure of 10 GPa. The temperature is controlled by a Langevin thermostat with damping constant $\gamma ^{-1}$=0.1~ps applied to the 4 atomic layers adjacent to the top and bottom rigid layers. All simulations are performed at room temperature (300~K). 

Randomly placed carbon atoms in the region between the two bare diamond slabs prevent cold welding, that is the joining of the two slabs~\cite{pastewka2010anisotropic}. When one or two graphene layers are present, these atoms allow  bond formation between the graphene layer and the diamond surface as we discuss later. 

It has been shown~\cite{pastewka2010anisotropic} that when two diamond slabs slide against each other, the crystalline structure at the interface is damaged, leading to an amorphous structure with a rate of amorphization which depends on the surface and on the sliding direction. 

\begin{figure}[htbp]
 \subfloat[]{\includegraphics[width=0.12\textwidth]{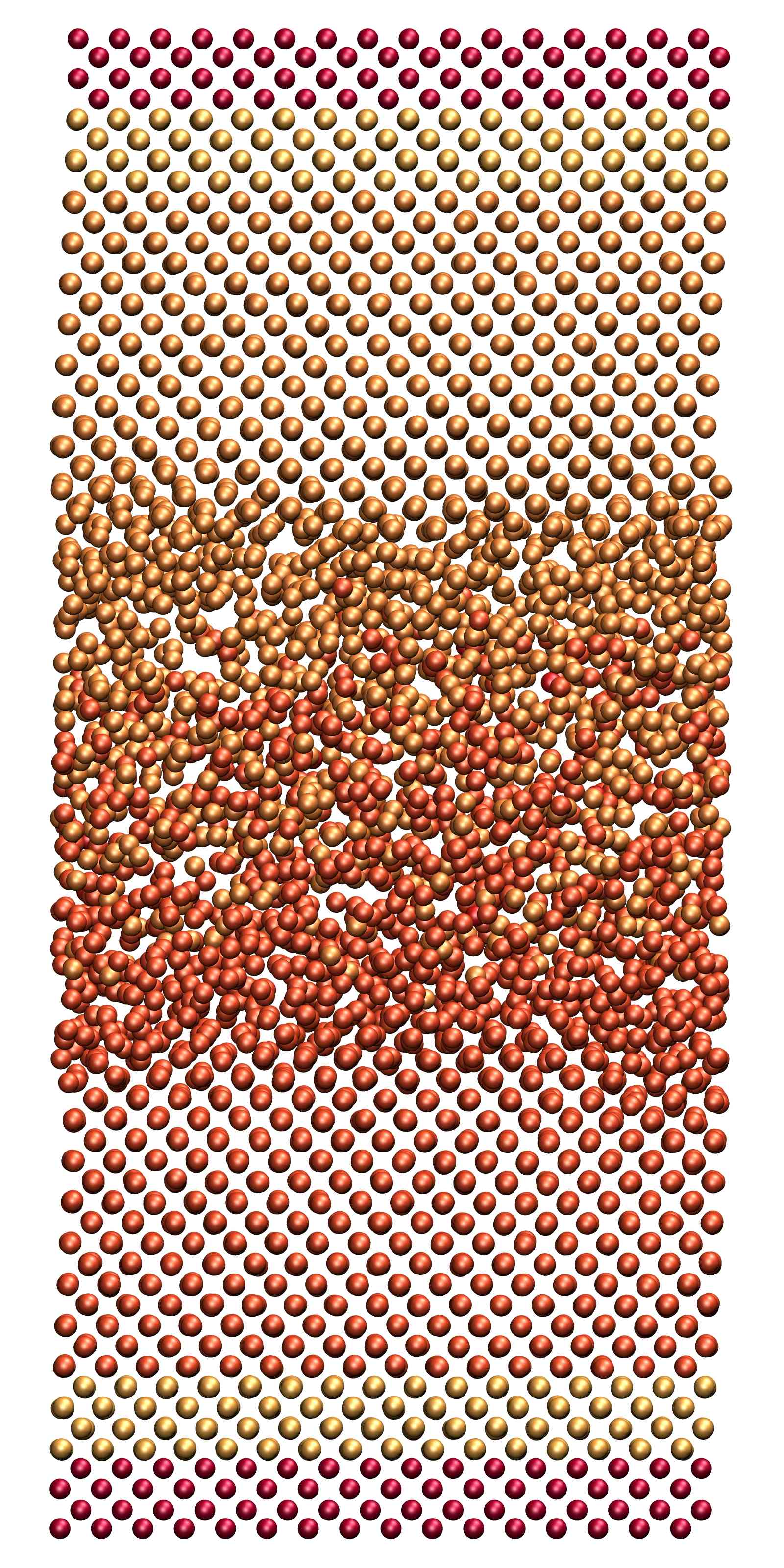}}
 \qquad
 \subfloat[]{\includegraphics[width=0.12\textwidth]{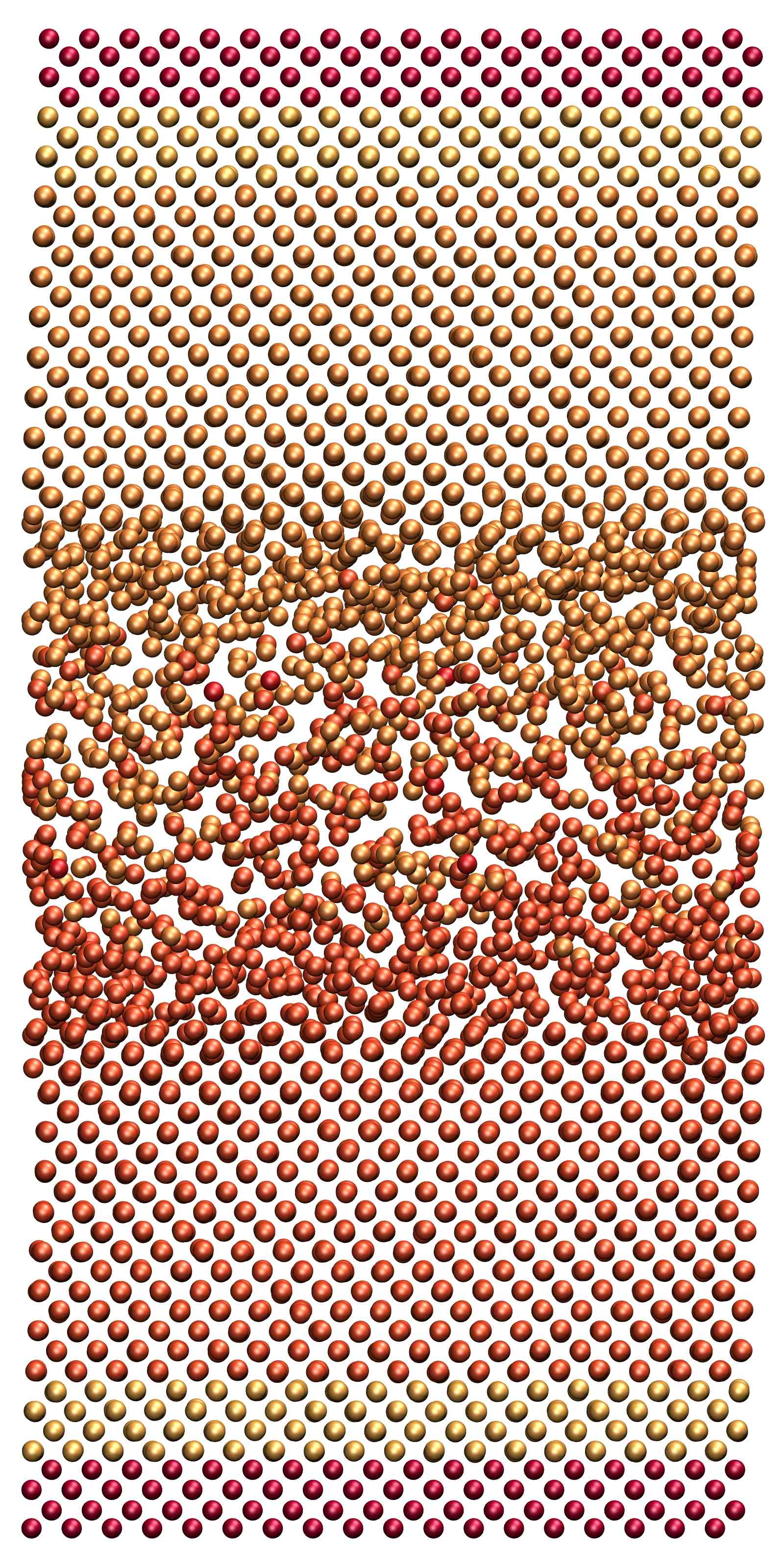}}
 \qquad
 \subfloat[]{\includegraphics[width=0.12\textwidth]{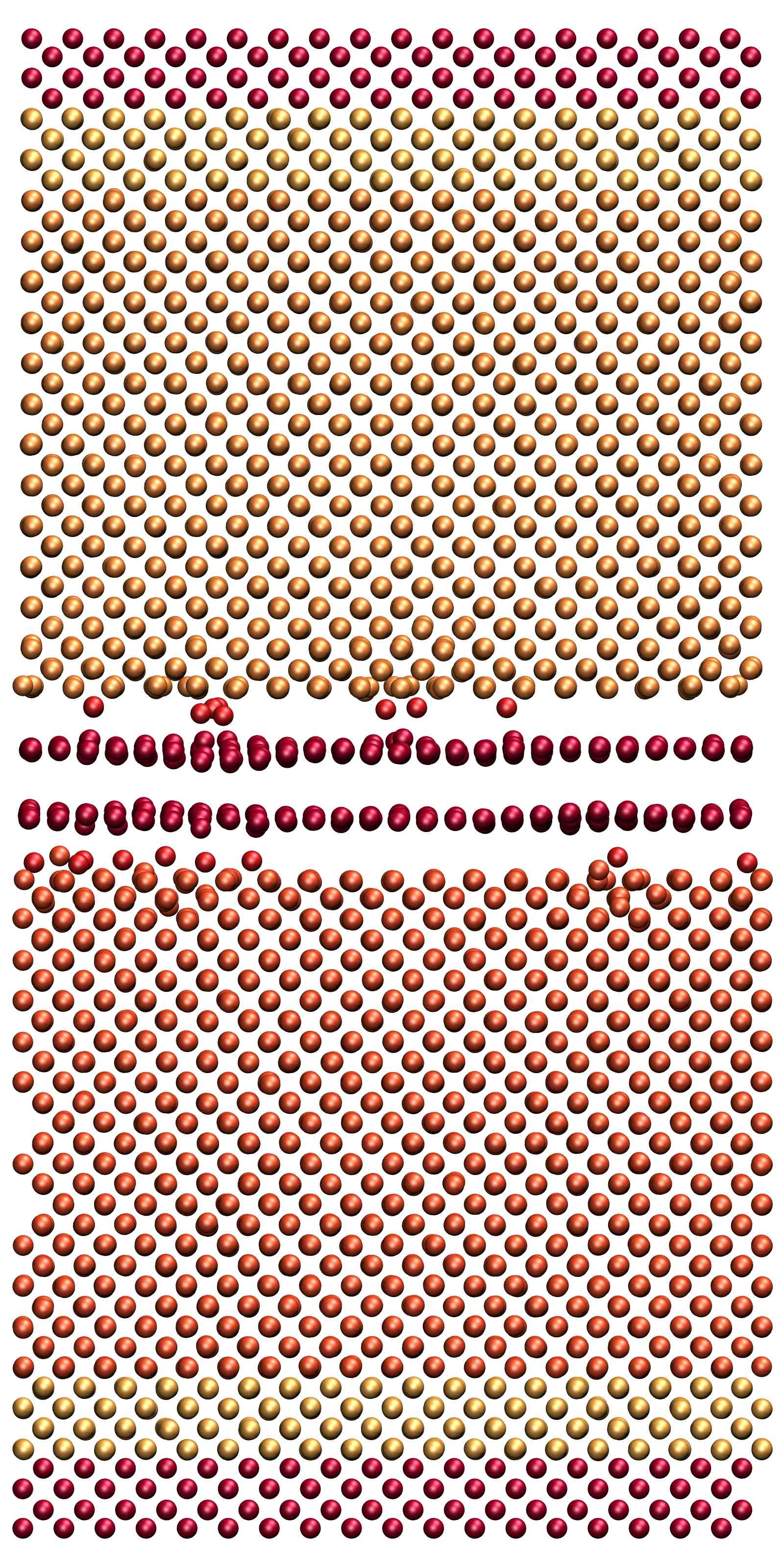}}
 \caption{(Color online) Structure of the samples of Fig.\ref{fig:Setup} after 2 ns of sliding: bare surfaces (a), with one layer of graphene (b) and with two layers of graphene (c). Notice that two layers of graphene prevent amorphization. }
 \label{fig:Wear}
\end{figure}

We consider $\left(100\right)$ diamond surfaces sliding in the $\left<100\right>$ direction, which is a fairly soft direction and find that the bare contact area transforms, as shown in Fig.~\ref{fig:Wear}a, into amorphous carbon with a $\sim 90$ \% sp$^2$ bonds. The precise percentage of  bonding in disordered, liquid or amorphous, phases may depend on the used potential~\cite{lucaliquid, pastewka2010atomistic}.

\begin{figure}[htbp]
\centering
 \subfloat[]{\includegraphics[width=0.45\linewidth]{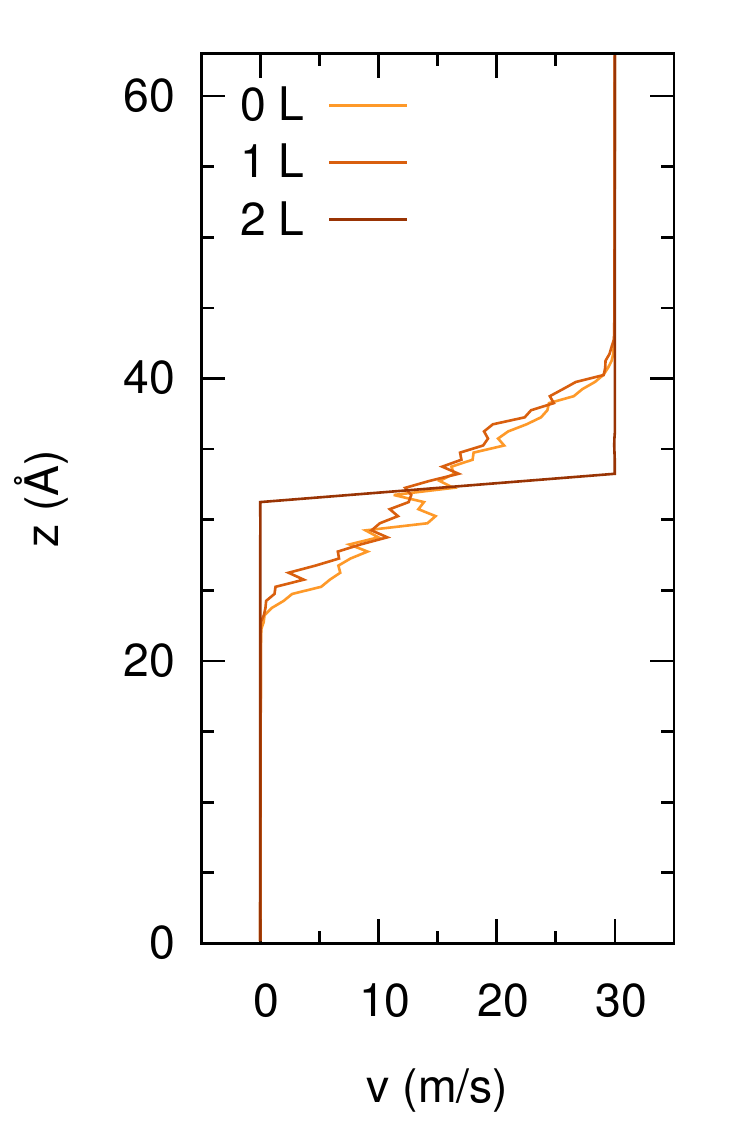}}
 \subfloat[]{\includegraphics[width=0.45\linewidth]{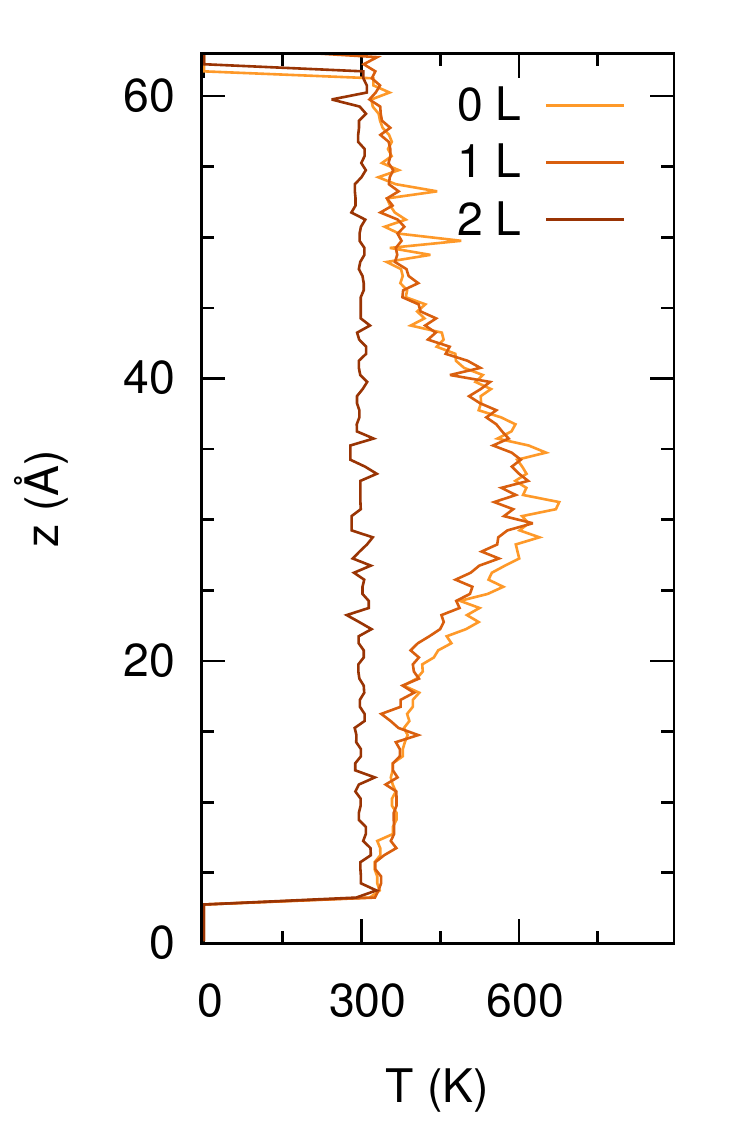}}
\caption{(Color online) a) Velocity in the sliding direction along the height of the sample after 2 ns for the samples of Fig.~\ref{fig:Setup} with zero (0~L), one (1~L) or two (2~L) graphene layers. For two graphene layers, a sharp transition is visible between the two slabs whereas the velocity gradually changes in for 0~L and 1~L due to the amorphous layer. b) Temperature along the height of the sample after 2 ns for samples with zero (0~L), one (1~L) or two (2~L) layers. The temperature is higher in the amorphous part.}
 \label{fig:velocity}
\end{figure}

A single graphene layer between the two surfaces leads to the same result, namely the graphene layer is destroyed within tens of picoseconds and the contact area becomes amorphous. This situation changes dramatically for a bilayer graphene (Fig.~\ref{fig:Wear}c). Sliding occurs in this case preserving the structure of the diamond surface as well as that of the bilayer. The different behaviors are also visible in the velocity and temperature profiles along the height of the sample, shown in Fig.~\ref{fig:velocity}. While the samples which degrade to amorphous carbon show a gradual change in velocity, the sample with two layers of graphene shows a sharp transition where the two slabs slide over each other. In this case, the temperature  remains constant at 300~K while for the amorphous contact area is raises to 600~K at the interface.

\begin{figure}[htbp]
 \includegraphics[width=0.8\linewidth]{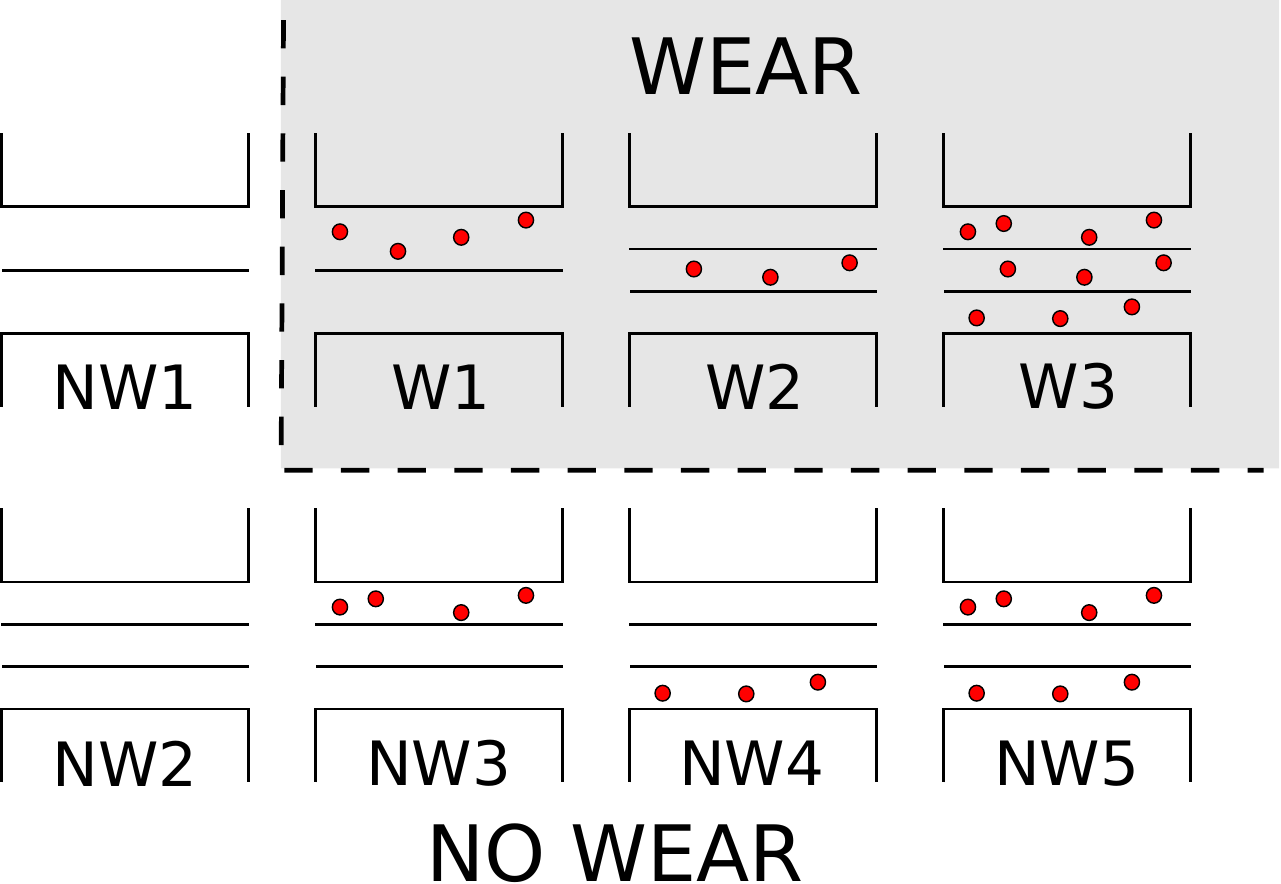}
 \caption{(Color online) Sketch of the simulated structures indicating the  graphene layers between the diamond surfaces and the presence of single carbon atoms. The shading is used to separate these structures into those that present wear after 2 ns and those that do not wear.}
 \label{fig:scheme}
\end{figure}

\begin{figure}[htbp]
 \subfloat[1 L]{\includegraphics[width=0.12\textwidth]{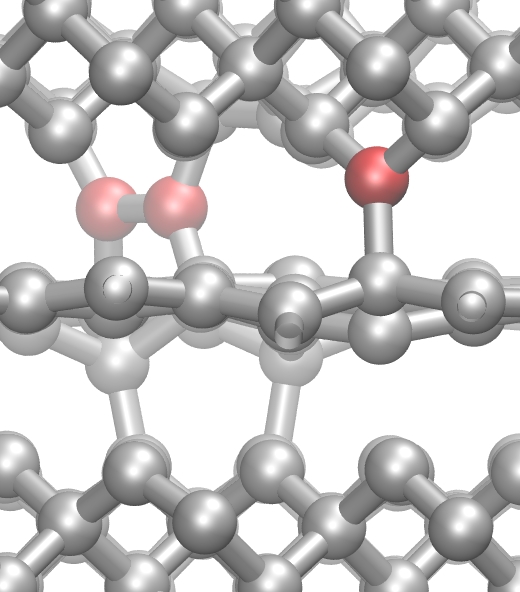}}
 \qquad
 \subfloat[2 L]{\includegraphics[width=0.12\textwidth]{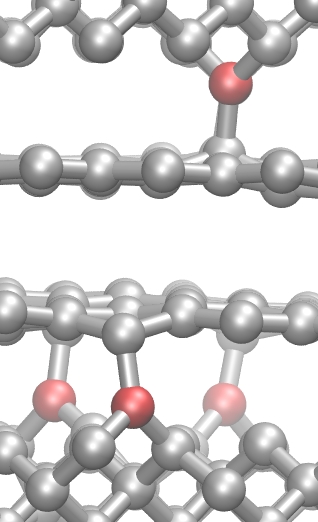}}
 \caption{(Color online) Snapshots after a few picoseconds of the structure with one  or two graphene layers. For the single graphene layer (a) bonds form on both sides and lead eventually to amorphization and wear. For the two graphene layers, instead,  the adatoms cannot induce bonds between the two graphene layers (b).}
 \label{fig:zoom}
\end{figure}

To understand the reason for the marked difference between one- and two-layer graphene coating of the diamond surface we have considered all the systems sketched in Fig.~\ref{fig:scheme} that we have divided into those that do not present wear within the timescale of our simulations and those that do. We see that it is important to consider the possibile imperfections of the surfaces or the presence of adsorbates and reactive molecules. In fact for ideally planar, clean surfaces with either one or two graphene layers in between but no adsorbates, no wear occurs during sliding (see panel NW1,NW2).  Adatoms in between the diamond surfaces and graphene lead to the formation of bonds as shown in Fig.~\ref{fig:zoom}, pulling graphene out of planarity. The consequences are very different for one or two layers. In fact, for one layer, once a bond is formed with the upper diamond surface, the deviations from planarity facilitate bonding of a neighboring atom  with the lower diamond surface. Bonds with upper and lower diamond surface become sp$^3$-like and propagate leading to an amorphous structure as in Fig.~\ref{fig:Wear}b. If instead there is a second layer of graphene, as in Fig.~\ref{fig:zoom}b, the bonding between the two graphene layers does not occur because it would require the two graphene layers to approach to distances below 2~\AA, which is prevented by the high energy barrier due to interlayer repulsion \cite{fahy1986pseudopotential,los2003intrinsic}. Wear occurs only when single carbon atoms are placed also between the graphene layers.

\begin{figure}[htbp]
 \subfloat[Grain boundary, top view]{\includegraphics[width=0.5\linewidth]{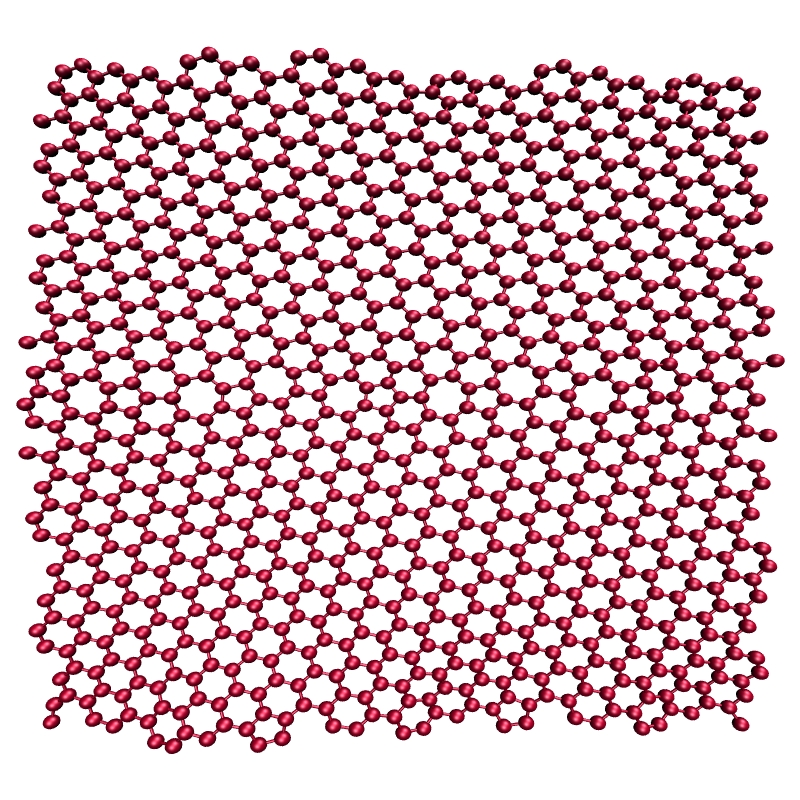}}
 \subfloat[Porous layer, top view]{\raisebox{10mm}{\includegraphics[width=0.4\linewidth]{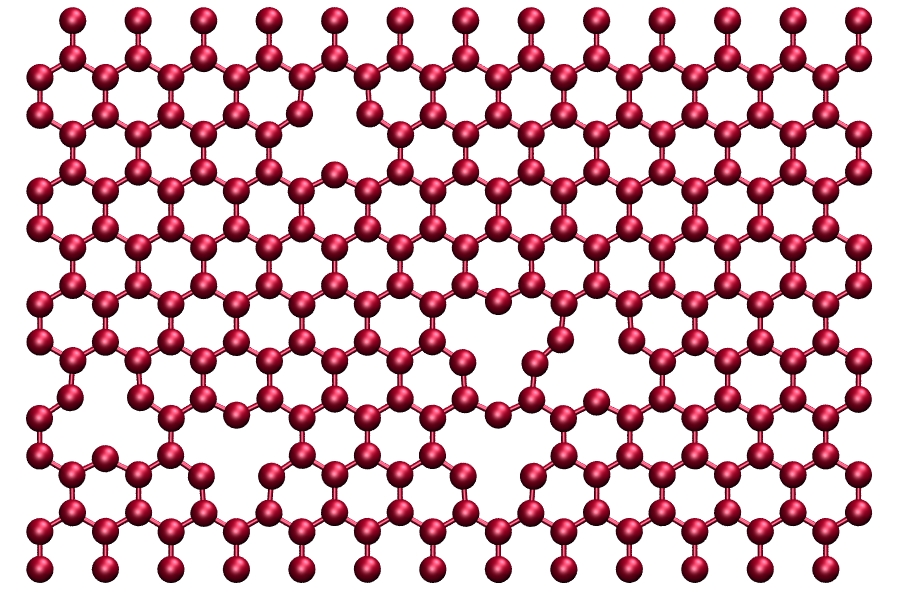}}}
 \caption{(Color online) (a) Graphene layer with a $\Sigma=39$ grain boundary~\cite{carlsson2011theory}. (b) Porous layer with vacancies obtained by randomly removing 3 \% of carbon atoms.}
 \label{fig:grainboundary}
\end{figure}

Next, we have considered less idealized systems by considering the most common defects in graphene, namely grain boundaries and vacancies. In the grain boundary shown in Fig.~\ref{fig:grainboundary}a the bonds form pentagons and heptagons which are more prone to rearrangement than the ideal hexagonal structure and the vacancies in Fig.~\ref{fig:grainboundary}b lead to unsaturated bonds. Also for these cases, 
we have found the same drastic difference between one and two layers. The only effect on the graphene layer with the grain boundary is a flattening of the curvature of the minimal energy structure without pressure~\cite{carlsson2011theory}. 
For the sample with vacancies,  we find that they remain intact and smooth when one percent or three percent of the atoms is missing. If we increase further the ratio of deleted atoms to five percent, the graphene layers degrade to amorphous carbon. That the graphene layers do not need to be perfect in order to inhibit wear is encouraging, since growth of perfect graphene is still a technological challenge.

As a last test, we have increased the potential energy corrugation which is underestimated by LCBOP. Therefore, we repeated the simulation of diamond with two graphene layers, but with the interactions between atoms in different graphene layers described by a registry-dependent potential~\cite{kolmogorov2000smoothest} but did not find any qualitative difference.  

In summary, we have shown that two layers of graphene between diamond slabs may provide a strong wear-resistant layer. While clean diamond surfaces or separated by only one layer of graphene transform to an amorphous phase during sliding under pressure, two layers of graphene preserve their structure and protect the diamond from wear. This result holds also when the graphene layers present defects such as a grain boundaries or vacancies. We believe that our findings can be relevant for the development of fully carbon based MEMS/NEMS.

We thank M. Patelkou for useful discussions. This work is part of the research program of the Foundation for Fundamental Research on Matter (FOM), which is part of the Netherlands Organisation for Scientific Research (NWO).

\end{document}